\documentclass[amssymb,reprint,twocolumn,pre]{revtex4}
\usepackage{graphicx}
\usepackage{psfrag}
\usepackage{color}
\usepackage{amsbsy}
\usepackage{amssymb}       

\def\dd{\mbox{d}}

\def\ve{\varepsilon}

\begin{document}



\title{The stochastic entry of enveloped
viruses: Fusion vs. endocytosis}

\author{Tom Chou \\ 
Depts. of Biomathematics \& Mathematics, UCLA, Los Angeles, CA 90095}


\date{\today}

\begin{abstract}
Viral infection requires the binding of receptors on the target cell
membrane to glycoproteins, or ``spikes,'' on the viral
membrane. The initial entry is usually classified as 
fusogenic or endocytotic. 
However, binding of viral spikes to cell surface receptors not only
initiates the viral adhesion and the wrapping process necessary for
internalization, but can simultaneously initiate direct fusion with
the cell membrane. Both fusion and internalization have been observed
to be viable pathways for many viruses. We develop a stochastic model
for viral entry that incorporates a competition between receptor
mediated fusion and endocytosis.  The relative probabilities of fusion
and endocytosis of a virus particle initially nonspecifically adsorbed
on the host cell membrane are computed as functions of receptor
concentration, binding strength, and number of spikes. We find
different parameter regimes where the entry pathway probabilities can
be analytically expressed. Experimental tests of our mechanistic
hypotheses are proposed and discussed.
\end{abstract}

\maketitle

%
%
%


\section{INTRODUCTION}

Viral entry mechanisms are typically classified as either endocytotic,
or as fusogenic \citep{REV0}.  In the latter, the virus membrane,
after association with the surface of the host cell, fuses and becomes
contiguous with the cell membrane. This process is mediated by the
binding of cell surface receptors to glycoprotein spikes on the viral
membrane surface, forming fusion competent complexes spanning the
viral and cell membranes. In endocytosis, the host cell first
internalizes the virus particle, wrapping it in a vesicle before
acidification-induced fusion with the endosomal membrane can occur.
Wrapping can occur only after cell surface receptors, that also act as
attachment factors, bind to the viral spikes. Experimentally, both
fusion with the cell membrane and internalization can be observed and
distinguished using microscopy \citep{MICRO0,MICRO1}.



 
Many viruses, such as influenza and hepatitis B, use endocytosis as
their primary mode of entry \cite{REV1,REV2}. The surface of an
influenza virus is coated with $\sim 400$ hemagglutinin (HA) protein
spikes \cite{400}. The HA adheres to sialic acid-containing
glycoproteins and lipids on the cell surface leading to wrapping of
the virus particle. Particle wrapping may also be mediated by the
recruitment of pit-forming clathrin/caveolin compounds
\cite{HIVENDO0}. Cell membrane pinch-off, leading to internalization
of the virus, usually requires additional enzymes such as dynamin and
endophilin \cite{DYNAMIN}. Endosomal acidification oligomerizes the
HA, priming them to fuse with the endosomal membrane.  Direct fusion
of the influenza virus with the host cell membrane is precluded since
HA is activated only in the acidic endosomal environment. However, low
pH conditions have also been shown to induce the direct fusion of
influenza virus with certain cells \cite{LOWPH}.

Recent experiments on the avian leukosis retrovirus (ALV) have
provided evidence both for a pH-dependent direct fusion mechanism
\cite{MELIKYAN,HARVARD}, and an endocytotic pathway
\cite{ANTIHARVARD}.  Moreover, the entry pathway of some viruses such
as Semliki Forest Virus (SFV) can be shifted from endocytosis to
fusion by acid treatment, but only in certain host cell types
\cite{SFV}. In this case, low pH triggering of receptor-primed
envelope glycoproteins can initiate fusion before the virus can be
wrapped and endocytosed. Vaccinia and HIV (typically infecting cells
via direct fusion) have also been shown to exploit both entry
mechanisms \cite{HIVENDO0,VACCINIA,HIVENDO1,KABAT0}.  For example,
fusion-independent mechanisms of HIV-1 capture and internalization in
mature dendritic cells, mediated by DC-SIGN \cite{DCSIGN0}, can be a
significant mode of HIV transmission through dendritic cells and
lymphatic tissue \cite{DCSIGN1}. Capture by DC-SIGN and CLEC-2
adhesion molecules also internalizes HIV in platelets
\cite{PLATELET}. Thus, depending upon physical conditions and cell
type, both entry pathways are potentially accessible to certain
viruses. The choice seems to depend on the type of receptors the
viruses engages, whether they are receptors/coreceptors that induce
fusion (perhaps triggered by low pH), or simply attachment factors
such as sialic acid-rich glycoproteins that do not induce fusion. In
this latter case, complete wrapping before an irreversible fusion
event is more likely to occur, and internalization is favored.

It is not surprising that subtle changes in the interactions
between viral membrane proteins and cell receptors dramatically
affect the infectivity of a virus, as recently demonstrated for the
1918 influenza virus \cite{INF}. In this paper, we model
virus-receptor kinetics and propose a mechanism consistent with the
above experimental observations, and that describes viral entry by
incorporating both fusion and endocytosis entry pathways in a
probabilistic manner.  In the next section, we develop a stochastic
one-species receptor model for the binding of receptors necessary to
start the virus wrapping process. These receptors, upon binding, can
induce membrane fusion at each receptor-spike complex.
In the Results, we find parameter regimes in which each of the entry
mechanisms dominate. Explicit expressions for the entry pathway
probabilities, as functions of the relevant kinetic rates, are given
in the Appendix. In the Discussion and Summary, we explore the
connection between our parameters and experimentally controllable
physical conditions such as receptor/coreceptor density, spike
density, and cell membrane rigidity. Experimental tests are proposed
and extensions of our analysis to more realistically incorporate
biological features are discussed.

\begin{figure*}[t!]
\begin{center}
\includegraphics[height=3.5in]{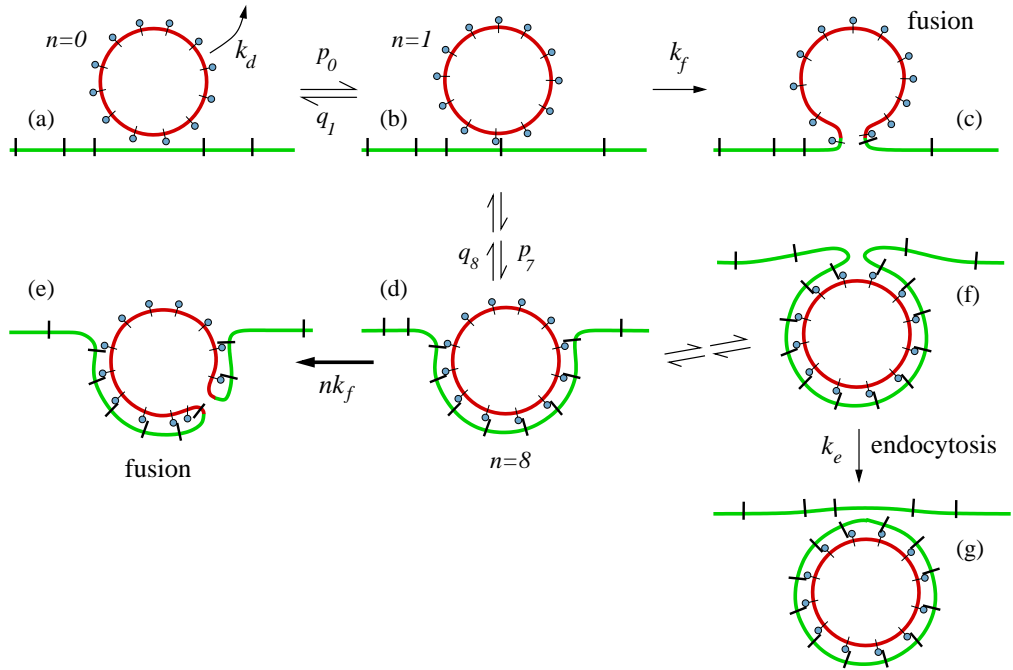}
\end{center} 
\vspace{-2mm}
\caption{A schematic of viral entry pathways. (a) A nonspecifically
  adsorbed virus particle can desorb with rate $k_{d}$, or it can (b)
  recruit and specifically bind a receptor. The receptor can
  immediately initiate membrane fusion with rate $k_{f}$ as shown in
  (c), or, it can recruit additional receptor molecules, inducing
  wrapping of the virus particle. From partially wrapped states (d),
  the virus can at any stage undergo membrane fusion (e), or, it can
  completely wrap and internalize the virus particle ((f) and (g)).}
\label{FIG1} 
\end{figure*}

\section{Single Receptor Kinetic Model}

The basic features of our proposed mechanism are shown in
Fig. \ref{FIG1}.  A virus particle initially nonspecifically adsorbed
on the cell membrane (without any bound receptors) can spontaneously
dissociate with rate $k_{d}$. Alternatively, mobile receptors in the
cell membrane can bind specifically to the glycoprotein spikes
(assumed here to be uniformly distributed) on the viral
surface. Successive addition of receptors to the viral ligands, when
$n$ are already attached, occurs with rate $p_{n}$.  Thus, the binding
of the first receptor occurs with rate $p_{0}$. Similarly, desorption
of the $n^{th}$ receptor occurs at rate $q_{n}$.  We consider the
adsorption of a single effective receptor or attachment factor to a
spike, lumping together the effects of multiple receptor/coreceptor
types. This approximation is valid when, for example, coreceptor
binding is highly cooperative such as suggested in the HIV infection
process where CD4 binding to the gp120 protein spike induces rapid
CCR5 coreceptor binding \cite{KABAT1}.

Receptors not only adhere the cell membrane to the viral membrane, but
can also initiate local membrane fusion at each receptor-spike complex
with rate $k_{f}$. Fusion can occur at any time during the receptor
recruitment process and is more likely to occur per unit time with
more bound receptors.  Receptor-spike complexes that are unable to
initiate fusion are described by a vanishing fusion rate
$k_{f}\rightarrow 0$.  However, if receptors have high fusogenicity
$k_{f}$, the virus might fuse only after a single receptor has attached.
Only if the system reaches a fully wrapped state with $N$ bound
receptors (Fig. \ref{FIG1}f), before any fusion event occurs, can
pinch-off and endocytosis occur with rate $k_{e}$. The number of
spikes $N$ is typically large, varying among viruses such as HIV
($N\sim 15$)\cite{SPIKESEM}, SIV ($N\sim 70$)\cite{SPIKESEM}, and
Influenza ($N\sim 400$)\cite{400}. The path to endocytosis is thus a
race between fusion and complete wrapping of a relatively 
large number $N$ of viral spikes.

\begin{figure}[h]
\begin{center}
\includegraphics[width=2.7in]{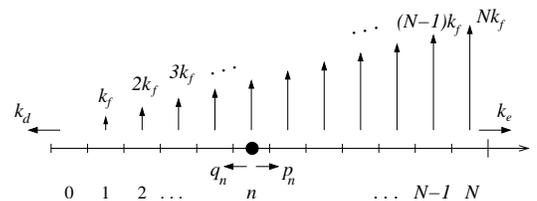}
\end{center} 
\vspace{-2mm}
\caption{The stochastic process representing the competition between
membrane fusion and endocytosis. The states $n$ correspond to the
number of receptor-spike complexes formed, while $N$ is the total
number of spikes on the virus membrane.  Each receptor-spike complex
can initiate membrane fusion with rate $k_{f}$. As more receptors are
bound, the total rate of fusion increases linearly.  The irreversible
pinch-off and endocytosis rate is denoted $k_{e}$.}
\label{KINETIC} 
\end{figure}

States in the model are labeled by the index $n$ (Fig. \ref{KINETIC}),
representing the number of formed receptor-spike complexes.  Starting
from a nonspecifically adsorbed virus particle denoted by state $n=0$,
the system progresses along the chain with the appropriate transition
rates corresponding to attachment and detachment of cell membrane
receptors. The probability $P_{n}(t)$ of having $n$ bound
receptors (or attachment factors) at time $t$ obeys the master equation


\begin{equation}
\begin{array}{l}
\displaystyle \dot{P}_{n}(t) = -(nk_{f}+p_{n}+q_{n})P_{n} + 
p_{n-1}P_{n-1} \\[12pt] 
\: \hspace{2.7cm} + q_{n+1}P_{n+1}, \quad 1\leq n \leq N-1, \\[13pt]
\displaystyle \dot{P}_{0}(t) = -(k_{d}+p_{0})P_{0}+q_{1}P_{1}, \\[13pt]
\displaystyle \dot{P}_{N}(t) = -(Nk_{f}+k_{e}+q_{N})P_{N}+p_{N-1}P_{N-1}.
\end{array}
\label{FP1}
\end{equation}

Although we describe the general viral entry process in terms of
recruitment and binding of a single type of receptor, our model
encompasses processes involving clathrin or calveolin aggregation and
pit formation, typically leading to endocytosis. All of these
biologically distinct, but physically similar mechanisms can be
analyzed by appropriately interpreting the rates. For example, the
nucleation of a clathrin-coated pit can be modeled by effective
binding and unbinding rates $p_{n}$ and $q_{n}$ that describe the
rates of clathrin addition and removal from the pit. Since individual
clathrin molecules are not known to induce membrane fusion, the fusion
rate $k_{f}=0$, and only endocytosis (or virus dissociation from the
cell surface) would occur. The fusion rate is also negligible if the
receptor is a simple attachment factor that adheres the membranes, but
does not facilitate fusion.

The receptor binding and unbinding rates, $p_{n}$ and $q_{n}$, are
related to the cell surface receptor density and the receptor-spike
binding strength, respectively. The Markov process shown in
Fig. \ref{KINETIC} implicitly assumes that the receptor recruitment is
not diffusion-limited -- the rate of addition of successive receptors
is independent of the history of previous receptor bindings.  

\begin{figure}[h]
\begin{center}
\includegraphics[width=2.8in]{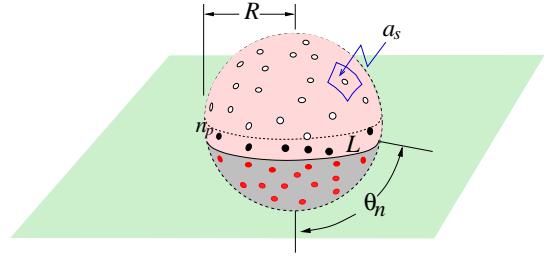}
\end{center} 
\vspace{-2mm}
\caption{Schematic of a partially wrapped virus particle. The unbound
  spikes above the contact region are represented by open circles,
  while the receptor-bound spikes in the contact region are
  represented by the red-filled circles. The number $n_{p}$ of spikes
  or spike-receptor complexes near the contact perimeter used to
  compute $p_{n}$ or $q_{n}$ via Eqs. \ref{PN} are shown as
  black dots.}
\label{WRAP} 
\end{figure}

Approximate forms for $p_{n}, q_{n}$ can be physically motivated by
considering the number of ways additional receptors can bind or
unbind, given that there already exist $n$ receptor-spike complexes.
As the membrane progressively wraps around the virus particle, the
rate of addition of the next receptor is proportional to the number
$n_{p}$ of unattached spikes bordering the virus-cell membrane contact
line $L$, as shown in Fig. \ref{WRAP}.

For a large number $N$ of uniformly distributed spikes on a virus
particle of radius $R$, the contact area $A_{c} \approx 4\pi R^{2}n/N
\approx 2\pi R^{2}(1-\cos \theta_{n})$ where $\theta_{n} =
\cos^{-1}\left[1-2n/N\right]$ is the angle subtended by the contact
line when $n$ receptors are attached. Since the area per spike is
$a_{s} \approx 4\pi R^{2}/N$, the number of spikes near the contact
perimeter $L= 2\pi R\sin\theta_{n}$ is found from $n_{p} \approx
L/\sqrt{a_{s}}$.
Upon using the explicit form for $\theta_{n}$, we find the number
$n_{p}$ of periphery spikes as a function of $n\geq 1$ receptor-bound
spikes, $n_{p} \sim \sqrt{\pi N} \sqrt{1-\left(1-2n/N\right)^{2}}$. At
a stage where $n\geq 1$ receptors have bound, approximately $n_{p}$
spikes are accessible for additional binding of receptors.  
Similarly, there are approximately 
$n_{p}$ receptor-spike complexes available for dissociation. 
Combinatorically, the receptor binding and unbinding rates take the form
\cite{LIGAND}

\begin{equation}
\begin{array}{l}
p_{n} \approx \displaystyle p_{0} {\sqrt{1-\left(1-{2n \over N}\right)^{2}} \over
\sqrt{1-\left(1-{2 \over N}\right)^{2}}}, \,\,\,
q_{n} \approx \displaystyle q_{1} {\sqrt{1-\left(1-{2n \over N}\right)^{2}} \over
\sqrt{1-\left(1-{2 \over N}\right)^{2}}},
\end{array}
\label{PN}
\end{equation}


\noindent where $1\leq n \leq N-1$, $p_{0}$ is the intrinsic rate of
binding the first receptor, and $q_{1}=q_{N}$ is the dissociation rate
of an individual receptor-spike complex.
Since $q_{1}$ is a spontaneous receptor-spike dissociation rate, it is
independent of receptor density and spike number.  If the
receptor-spike binding energy is at least a few $k_{B}T$, we also
expect $q_{n}$ to be relatively insensitive to the cell membrane
bending rigidity.

A number of physical attributes and biological intermediates can be
incorporated into the rate parameters to address more complicated
microscopic processes. For example, if thermal fluctuations are
rate-limiting, there would be an additional factor in $p_{n}$
reflecting the probability per unit time that a patch of membrane
fluctuates to within a distance of the virus surface spike that allows
receptor-spike binding. The dynamics of this process depends on the
cell membrane tension and rigidity, the typical spike spacing, and
potentially the viscosity of the extracellular environment. For stiff
membranes under tension, the wrapping of spherical particles
encounters an energy barrier near half-wrapping \cite{DESERNO}, which
can be incorporated into the dynamics by assuming an additional factor
in $p_{n}$ that has a minimum near $n\approx N/2$. Thus, the energy
barrier associated with membrane bending will tend to flatten the
$n$-dependence of $p_{n}$.

Particular aspects of the entry process can also influence estimates
of the other rates.  If the viral spikes are mobile and can aggregate
to the initial focal point of adhesion, the target cell membrane is not able to
fully wrap and endocytosis is prevented. The overall kinetic scheme
remains unchanged except with $k_{e} \approx 0$. Finally, the
recruitment of secondary coreceptors that occurs in, {\it e.g.} HIV
fusion, can also be developed within our current framework and is
discussed in the Discussion and Summary section. Although some of the
physical details described above influence the specific values of
$p_{n}$, we will show that for large $N$, the qualitative behavior of
our model can be summarized by distinct parameter regimes, somewhat
insensitive to the precise values of $p_{n}, q_{n}$.



\section{Results}

\begin{figure}
\begin{center}
\includegraphics[width=3.4in]{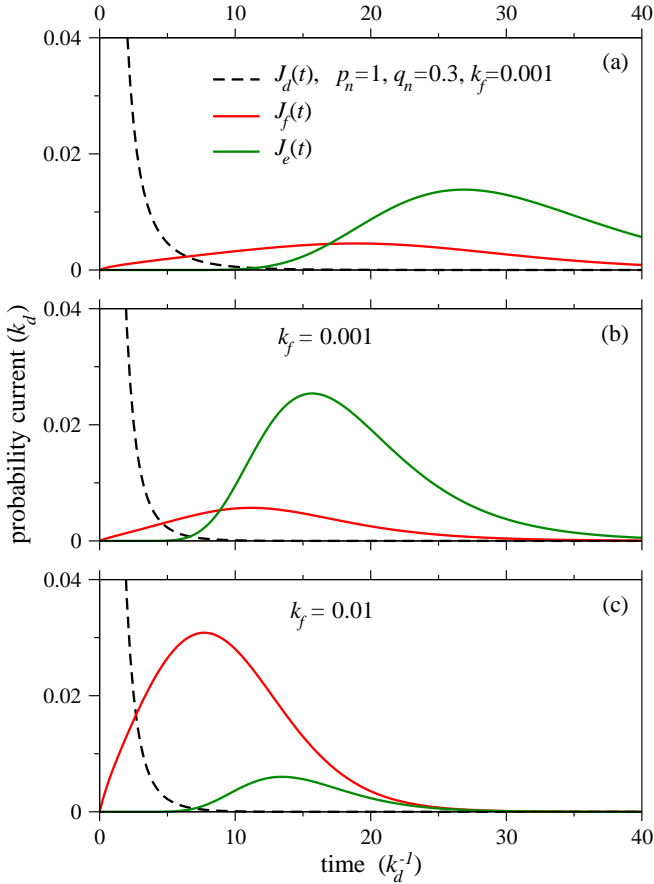}
\end{center} 
\caption{The currents through each pathway for $N=20$ spikes. The dashed black
curve is the current for desorption, while the red and green curves
are the currents for fusion and endocytosis, respectively.  Time is
measured in units of $k_{d}^{-1}$, and all rates are normalized with
respect to $k_{d}$. The parameters used in all plots are $p_{0}=1$,
$q_{1}= 0.3$, and $k_{e} = 0.3$.  (a) The currents for constant
$p_{n}=1$, $q_{n}=0.3$ and fusion rate $k_{f} = 0.001$. (b) The same
parameters except that Eqs. \ref{PN} are used for the rates $p_{n},
q_{n}$. (c) The currents with $p_{n}, q_{n}$ as in (b), except that
the fusion rate of each spike-receptor complex is increased to
$k_{f}=0.01$.}
\label{CURRENTS} 
\end{figure}

We solved Eq. \ref{FP1} for the probabilities $P_{n}(t)$ given an
initial condition $P_{n}(t=0) = \delta_{n,0}$. Using these
probabilities, we find the probability currents through the
dissociation, fusion, and endocytosis pathways, $J_{d}(t) =
k_{d}P_{0}(t)$, $J_{f}(t) = k_{f}\sum_{n=1}^{N} nP_{n}(t)$, and
$J_{e}(t) = k_{e}P_{N}(t)$, respectively. Figure \ref{CURRENTS} shows
the current for each pathway as a function of time (in units of
$k_{d}^{-1}$). Note that detachment, fusion, and endocytosis arise
sequentially in time. 
Upon comparing Fig \ref{CURRENTS}(a) and (b), we
see that changing $p_{n}, q_{n}$ from constant values to the forms in
Eqs. \ref{PN} shifts the currents through the fusion and endocytotic
pathways to earlier times.  This speed-up is simply a consequence of
the larger hopping rates $p_{n}, q_{n}$, especially for $n\approx
N/2$. Nonetheless, the specific form of $p_{n}, q_{n}$, provided they
are slowly varying in $n$, only quantitatively affect the timing of
the onset of the currents.  The dramatic variations in the infection
pathway taken come with changes in $k_{f}$.  When the fusion rate
$k_{f}$ is increased, endocytosis is suppressed in favor of fusion as
shown by Figs. \ref{CURRENTS}(b) and \ref{CURRENTS}(c).

\begin{figure}
\begin{center}
\includegraphics[width=3.4in]{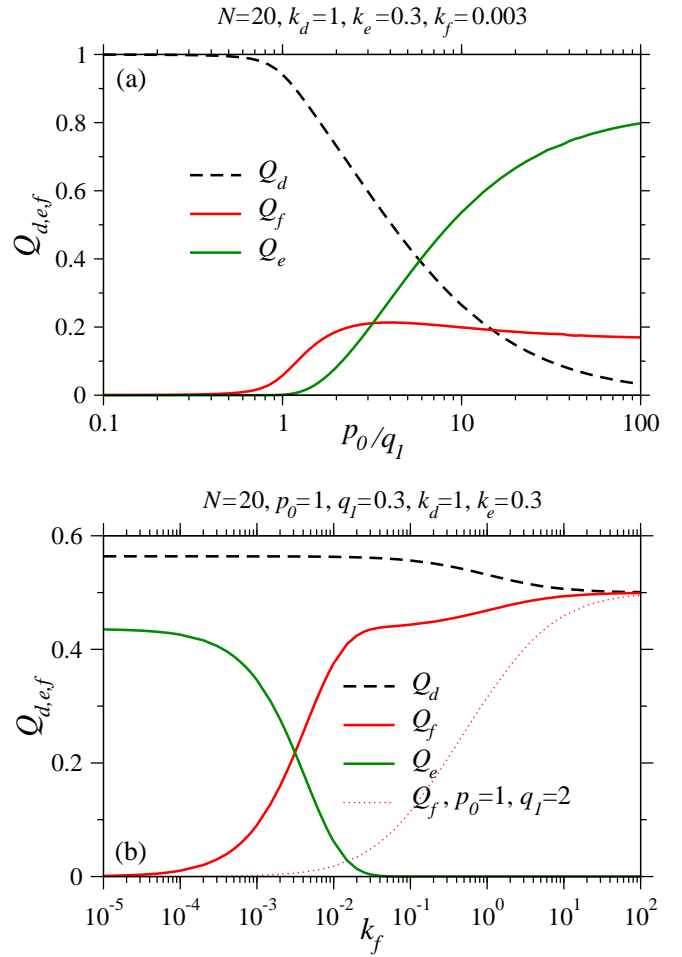}
\end{center} 
\caption{Numerical solutions of entry probabilities $Q_{i}$ (all rates
normalized by $k_{d}$). (a) The entry probabilities as a function of
$p_{0}$. Endocytosis arises only for larger $p_{0}>q_{1}$, after the
fusion probability becomes significant. Parameters used are $N=20,
k_{f} = 0.003,k_{e}=0.3$ (all rates are normalized by $k_{d}$). (b)
The probabilities of dissociation (dashed), fusion (red), and
endocytosis (green) as functions of the individual receptor-spike
fusion rate $k_{f}$.  Here, $q_{1}=0.3$. The thin dashed red curve
corresponds to a faster receptor detachment rate ($q_{1}=2$) which
prevents endocytosis.}
\label{FRACTION} 
\end{figure}

Upon time-integrating the currents, we find the total probabilities
$Q_{i}=\int_{0}^{\infty} J_{i}(t)\dd t, \, (i = d,e,f)$ for each
pathway. Note that from probability conservation,
$Q_{d}+Q_{e}+Q_{f}=1$.  In Figure \ref{FRACTION}, we use the binding
and unbinding rates given by Eqs. \ref{PN} to numerically compute the
entry pathway probabilities $Q_{i}$.  
Fig. \ref{FRACTION}(a) shows the
pathway probabilities as a function of the intrinsic receptor binding
rate $p_{0}/q_{1}$. This ratio is a measure of the density-dependent
free energy $\Delta G$ of the spike-receptor binding: $p_{0}/q_{1}
\sim e^{\Delta G/kT}$ \cite{LIGAND}. In the simplest limit of extremely
low receptor density, ($p_{0} \rightarrow 0$), $Q_{d} \sim 1-{\cal
O}(p_{0}), Q_{f} \sim {\cal O}(k_{f}p_{0}),$ and $Q_{e} \sim 0$, and
only dissociation can occur. As $p_{0}$ is increased, the probability
of fusion increases at the expense of desorption. Endocytosis remains
negligible as long the states that occur with any appreciable
probability are those with small $n$. Only as $p_{0}\gg q_{1}$ does
the probability of endocytosis become appreciable and approach the
asymptotic expression given by Eq. \ref{q=0} in the Appendix.

Figure \ref{FRACTION}(b) shows the total probabilities $Q_{d}$ of
dissociation, $Q_{f}$ of fusion, and $Q_{e}$ of endocytosis as
functions of the fusion rate $k_{f}$.  For large detachment rates ({\it e.g.}
$q_{1}=2> p_{0}=1$), $Q_{e}\approx 0$, and fusion can occur only at
large $k_{f}$, as shown by the thin dotted curve.  For the parameters
used, $N=20, p_{0}=1, q_{1}=k_{e}=0.3$ (normalized by $k_{d}$), the
transition from a predominantly endocytotic pathway to a predominantly
fusion pathway occurs for $10^{-3}\lesssim k_{f} \lesssim
10^{-2}$. When $k_{f} \gg 10^{-2}$, the sum of fusion probabilities
over all intermediate states is appreciable, preventing endocytosis.
Therefore, only for a particularly small fusion rate $k_{f}$, and
nonnegligible endocytosis rate $k_{e}$ is internalization possible.
We show in the Appendix that generally, if $p_{n} > q_{n}$, the
effective drift of the stochastic system towards the wrapped state
renders the partitioning between fusion and endocytosis controlled by
the fusion rate $k_{f}$. If this is the case, the transition from
endocytosis to fusion occurs at approximately $k_{f} \sim
(p_{n}+q_{n})/N^{2}$, as is confirmed by the numerical solution for
$N=20$ shown in Fig. \ref{FRACTION}(b).

\begin{figure}
\begin{center}
\includegraphics[width=3.4in]{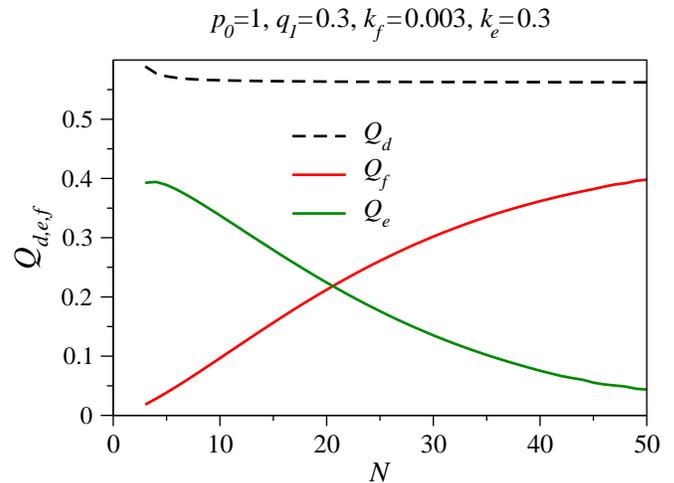}
\end{center} 
\caption{The pathway dependence on receptor association/dissociation rates and
the number $N$ of virus spikes. The number of spikes controls which
regime Eq. \ref{ASYMP0} or Eq. \ref{ASYMP2} is valid.  Large $N$
enhances fusion almost entirely at the expense of endocytosis.}
\label{FRACTION_N} 
\end{figure}

Although we have chosen $N=20$ as a repreentative spike number,
different viruses and their variants can have a widely varying number
of active spikes. In Fig. \ref{FRACTION_N} we show how the entry
pathway probabilities depend on the number $N$ of active viral spikes.
As $N$ is increased, the probability for fusion increases at the
expense of endocytosis. For $N\rightarrow \infty$, and a nonzero
$k_{f}$, $Q_{e}\rightarrow 0$ since fusion will likely occur during
the infinitely long wrapping processes. For small $N$, the plotted
probabilities must be interpreted with an $N$-dependent $k_{e}$.
Suppose $N \lesssim 10$. Even if all spikes are receptor-bound, the
membrane has an appreciable distance to bend and before full wrapping
and endocytosis can occur. Effectively, the fusion rate $k_{e}$ starts
to decrease if $N$ gets small such that the rate of membrane
fluctuations over a typical interspike distance decreases.

Provided $p_{n} - q_{n} \gg 1/N$, asymptotic analysis of the solutions
reveal qualitatively different behaviors depending upon how the fusion
rate $k_{f}$ compares with $1/N$. If the receptor-spike complex is
highly fusion competent such that $k_{f}/(p_{n}+q_{n}) \gg 1/N$, the
probability of reaching a completely wrapped ($n \approx
N$) state is exponentially small and endocytosis cannot occur.  Here,
the virus pathway is nearly entirely partitioned between dissociation
and fusion, as indicated by the asymptotic expressions for $Q_{i}$
given by Eqs. \ref{ASYMP0} in the Appendix.

If $1/N^{2} \ll k_{f}/(p_{n}+q_{n}) \ll 1/N$ , the receptor-spike
complex has intermediate fusogenicity. In this case, the
time-integrated probability $\int_{0}^{\infty} P_{n}(t)\dd t$ used to
construct $Q_{i}$ remains small for $n\approx N$ and the probability
of endocytosis is still exponentially small, despite the smaller
$k_{f}$.  In this regime, we find in the Appendix (Eq. \ref{ASYMP2}),
an expression for $Q_{d}$. Only when $k_{f}/(p_{n}+q_{n}) \ll 1/N^{2}$
is virus internalization appreciable. Expressions for $Q_{f}$ and
$Q_{e}$ in this very weak fusogenicity limit are also displayed in
Eq. \ref{ASYMP2} of the Appendix. The expression we find 
agree with the exact numerical evaluation of $Q_{d,e,f}$ 
from solving Eq. \ref{FP1}.

\section{Discussion and Summary}

Fusion can be directly distinguished from endocytosis by imaging
fluorescent markers loaded into viruses or model vesicles.  Upon
fusion, one expects to see an immediate release of marker into the
periphery of the target cell. Similarly, single-liposome fluorescence
imaging experiments that label and detect vesicle lipids and as they
mix with a supported bilayer upon fusion \cite{UIUC}, can be used as
{\it in vitro} model systems for virus fusion.  In such experiments,
the model parameters $p_{n}, q_{n}, k_{f}, k_{e}, k_{d}$ can be tuned
by controlling certain physical chemical properties, enabling one to
dissect the mechanism of viral entry.  The entry pathways delineated
by the different parameter regimes described in the previous
section, and by the asymptotic formulae given in the Appendix, provide
a framework for analyzing and designing viral entry experiments.

The receptor density plays the first critical role via the binding
rate parameter $p_{n}$.  For low receptor densities, and
proportionately lower $p_{n}$ (but fixed spontaneous detachment rate
$q_{n}$), the virus particle can only dissociate or fuse. Although
lowering receptor concentration decreases the overall entry
probability, it can increase $Q_{f}/Q_{e}$, the fusion probability
relative to endocytosis probability (see Fig. \ref{FRACTION}(a)). Only
for $p_{n}-q_{n} > O(1/N)$ can the receptor spike complex fusion rate
$k_{f}$ become important in determining whether fusion or endocytosis
occurs.  In order for endocytosis to occur, the fusion rates must be
small such that $k_{f}/(p_{n}+q_{n})\ll 1/N^{2}$.
The rate $p_{n}$ can also be substantially decreased by
increasing the target membrane surface tension, thereby suppressing
the thermal fluctuations of the membrane required to bring cell
receptors and viral spikes into proximity.

The rapid drop-off in endocytosis predicted as the fusion rate is
increased from $k_{f}/(p_{n}+q_{n})\ll 1/N^{2}$ to
$k_{f}/(p_{n}+q_{n})\gg 1/N^{2}$, especially for large $N$, shows that
tuning physical conditions (such as pH or temperature) that affect the
fusogenicity of receptor-spike complexes, $k_{f}$, can have a large
effect on the viral entry pathway. Recent experiments by Melikyan {\it
  et al.}  \cite{TAS0}, Henderson and Hope \cite{TAS1}, and others
have shown a rate-limiting intermediate in the HIV fusion process that
can be arrested by lowering temperature.  Since CD4 binding was not
the rate limiting step, lowering the temperature decreases $k_{f}$ to
a degree presumably much less that $(p_{n}+q_{n})/N^{2}$, preventing
fusion.  If these systems have the necessary endocytotic machinery and
support pinch-off, lowering temperature and arresting the
receptor-spike fusion complex while retaining the adhesive wrapping of
receptor-spike binding would enhance the endocytotic pathway. The
effective rate $k_{f}$ can also be lowered by cross-linking (with {\it
  e.g.}, defensins) membrane glycoproteins, rendering their complexes
with viral spikes fusion incompetent \cite{DEFENSIN}. However, if the
cross-linked glycoproteins retain their attraction for the viral
surface, the probability of wrapping and internalization would
increase.  If an independent measurement or estimate of $p_{n}+q_{n}$
is available, the dependence of $k_{f}$ on temperature, pH, and
chemical modification can be probed.

Finally, pathways to fusion and endocytosis can diverge for systems
that require both primary receptors and secondary receptors
(coreceptors).  If two species are required, one for adhering cell and
viral membranes, and another to induce fusion, endocytosis will be
favored, all else being equal.  In this case, the initial receptor
binding only causes the cell membrane to wrap around the virus
particle. An additional coreceptor must diffuse and bind to the
spike-receptor adhesion complex to induce fusion. A highly cooperative
receptor-coreceptor interaction, such as in HIV fusion involving CD4
and CCR5/CXCR4, is still modeled by Eq. \ref{FP1}, but with the
binding rates $p_{n}$ interpreted as an effective binding rate for
both classes of receptor, proportional to the product of their surface
concentrations. However, if the coreceptor density and/or mobility is
limiting \cite{KABAT0}, the binding of receptors occur first, with
coreceptor priming and formation of a fusion competent receptor-spike
complex occurring slowly.  This allows time for receptor (adhesion
molecule) mediated membrane wrapping of the entire virus, enhancing
the likelihood of endocytosis. Thus, by maintaining a high adhesion
receptor density, and lowering the fusion-enabling coreceptor density,
one enhances the endocytotic pathway.


%
%
%
%

\section{Appendix}

Consider the Master Equation \ref{FP1} in the form $\dot{{\bf P}}(t) =
-\left[{\bf M}_{0}+{\bf M}_{f}\right]{\bf P}$, where ${\bf M}_{0}$ is
the conserved random walk transition matrix involving only $k_{d},
p_{n}, q_{n}$, and $k_{e}$, and ${\bf M}_{f} = k_{f}\mbox{diag}(n)$ is
decay term arising from fusion. Large $N$
expressions for the entry pathway probabilities $Q_{i}$ can 
be obtained in different limits. 



In the limit where the receptor binding is irreversible ($p_{n}/q_{n}
\rightarrow \infty$), Eq.\, \ref{FP1} can be solved exactly:
\begin{equation}
\begin{array}{l}
\displaystyle Q_{d} =  {k_{d} \over p_{0}+k_{d}} \\[13pt]
\displaystyle Q_{f} = {1\over p_{0}+k_{d}}\left[p_{0}-{k_{e}p_{N-1}\over k_{e}+Nk_{f}}
\prod_{m=1}^{N-1} {p_{m-1} \over p_{m}+mk_{f}}\right] \\[13pt]
\displaystyle Q_{e} = {k_{e} \over k_{e}+Nk_{f}}{p_{N-1}\over p_{0}+k_{d}}
\prod_{m=1}^{N-1} {p_{m-1} \over 
p_{m} + mk_{f}}.
\label{q=0}
\end{array}
\end{equation}

\noindent If $q_{n}=0$, the probability of dissociation is fixed by
$k_{d}$ and $p_{0}$, and the remaining current is partitioned between
fusion and endocytosis.  For a given $p_{0}, k_{f}, k_{e}$, this limit
gives the highest probability of the maximally-wrapped state and the
highest endocytosis probability.

When $p_{n}-q_{n} \gg 1/N$, there are two possible limits
corresponding to high receptor-spike fusion rates,
$k_{f}/(p_{n}+q_{n})\gg 1/N$, and intermediate fusion rates $1/N \gg
k_{f}/(p_{n}+q_{n}) \gg 1/N^{2}$.  For $k_{f}/(p_{n}+q_{n})\gg 1/N$,
$\int_{0}^{\infty}P_{n}(t)\dd t$ is nonnegligible only for $n\approx
0$ and $\int_{0}^{\infty}P_{N}(t)\dd t$ is exponentially small.  A
small $n$ local analysis of Eq. \ref{FP1} yields

\begin{equation}
\begin{array}{l}
\displaystyle Q_{d} \approx {k_{d}\over k_{d}+p_{0} - p_{0}q_{1}/(p_{1}+q_{1})} \\[13pt] 
\displaystyle Q_{f} \approx {p_{0} \over k_{d} +p_{0} + k_{d}q_{1}/p_{1}}, 
\quad \quad \,\,\,\, \mbox{and} \\[13pt] 
\displaystyle Q_{e}
\approx 0, \qquad \qquad \qquad \qquad \qquad \mbox{for} \quad {k_{f}\over p_{n} +
q_{n}} \gg 1/N,
\end{array}
\label{ASYMP0}
\end{equation}

\noindent explicitly showing the absence of endocytosis. The values
for $Q_{d}$ and $Q_{f}$ corresponding to the parameter values used in
Fig. \ref{FRACTION_N} are $Q_{d} \approx 0.56$ and $Q_{f}\approx
0.435$, which agree well with the large $N$ limit of the numerical
soution.

In an intermediate fusogenicity regime, $1/N \gg k_{f}/(p_{n}+q_{n})
\gg 1/N^{2}$, $\int_{0}^{\infty}P_{n}(t)\dd t$ is nearly constant for
$n\ll N$.  However, $\int_{0}^{\infty}P_{N}(t)\dd t$ remains small and
endocytosis, although still unlikely, occurs with slightly higher
probability than in the high fusogenicity limit. In this intermediate
limit, one might first attempt perturbation theory about $k_{f}=0$
(since its largest element of ${\bf M}_{f}$, $Nk_{f} \ll p_{n}+q_{n}$,
is much small than the typical elements of ${\bf M}_{0}$).  However,
as $1/N$ decreases, so do the spacings between eigenvalues of the
zero-fusion transition matrix ${\bf M}_{0}$. Heuristically, for
perturbation theory to be accurate for slowly varying $p_{N}, q_{N}$,
the largest of the diagonal correction terms, $Nk_{f}$, must be
smaller than $(p_{n}+q_{n})/N$, the typical spacing between
eigenvalues. First order perturbation for {\it all} $P_{n}(t)$ is
accurate only if $k_{f}/(p_{n}+q_{n}) \ll 1/N^{2}$, as is explicitly
shown by Eqs. \ref{q=0}. Upon expanding $Q_{f}$ and $Q_{e}$ (from
Eqs. \ref{q=0}) in powers of $k_{f}$, one finds ${\cal O}(N^{2})$
corrections terms.  Thus, the expansion is only accurate if
$k_{f}/(p_{n}+q_{n})\ll 1/N^{2}$. Endocytosis is preempted by fusion
or dissociation only when perturbation theory about a nonnegligible
$Q_{e}$ fails (when $k_{f}/(p_{n}+q_{n})\not\ll 1/N^{2}$).
Perturbation results for the few-receptor states $n\approx 0$
important for the dissociation probability $Q_{d}$, are still valid as
long as $k_{f}/(p_{n}+q_{n}) \ll 1/N$. However, the perturbation
results that include the flux through larger ($n\approx N$) states are
accurate only if the receptor-spike complexes are extremely
inefficient at initiating membrane fusion, and $k_{f}/(p_{n}+q_{n})\ll
1/N^{2}$.  Thus, for small enough $k_{f}/(p_{n}+q_{n})$, we find

\begin{widetext}
\begin{equation}
\begin{array}{ll}
\displaystyle Q_{d} \approx
{k_{d}+k_{d}\sum_{n=1}^{N-1}\prod_{i=1}^{n}(q_{i}/p_{i}) + (k_{d}
q_{N}/k_{e})\prod_{i=1}^{N-1}(q_{i}/p_{i})\over k_{d}+p_{0} +
k_{d}\sum_{n=1}^{N-1}\prod_{i=1}^{n}(q_{i}/p_{i}) + (k_{d}
q_{N}/k_{e})\prod_{i=1}^{N-1}(q_{i}/p_{i})}, &  k_{f}/(p_{n}+q_{n}) \ll N^{-1} \\[13pt] 
\displaystyle Q_{f} \approx {p_{0} - q_{1} \over k_{d}+p_{0}-q_{1}} - Q_{e}, &
k_{f}/(p_{n}+q_{n})\ll N^{-2}\quad 
\mbox{where} \\[13pt] 
\displaystyle Q_{e} \approx {k_{e} \over
k_{e}+p_{0}^{-1}k_{d}k_{e}\left[1+\sum_{j=1}^{N-1}\prod_{i=1}^{j}\left({q_{i}\over
p_{i}}\right)\right] + k_{d}\prod_{j=0}^{N-1}\left({q_{j+1}\over
p_{j}}\right)}, & k_{f}/(p_{n}+q_{n})\ll N^{-2}, 
\end{array}
\label{ASYMP2}
\end{equation}
\end{widetext}

\noindent {\it independent} of $k_{f}$. Equations
\ref{q=0}-\ref{ASYMP2} give estimates for the entry probabilities
$Q_{d,e,f}$ in the different parameter regimes. For small $N$ such that
$k_{f}/(p_{n}+q_{n})\ll N^{-2}$, $Q_{e}\approx 0.411$, which agrees
well with the limit shown in Fig. \ref{FRACTION_N}.

\vspace{3mm}

The author thanks B. Lee, G. Lakatos, and
M. D'Orsogna for valuable discussions. This work was supported by the
NSF (DMS-0349195) and by the NIH (K25AI058672).


\vspace{1mm}

\begin{thebibliography}{99}

\bibitem{REV0} Barocchi MA, Masignani V, Rappuoli R (2005) 
{\it Nature Reviews Microbiology} 3:349-358.

\bibitem{MICRO0} Lakadamyali M, Rust MJ, Babcock HP,
Zhuang X (2003) 
{\it PNAS} 100:9280-9285.

\bibitem{MICRO1} Markosyan RM, Cohen FS, Melikyan GB (2005)
{\it Mol. Biol. Cell.} 16:5502-5513.

\bibitem{REV1} Cross KJ, Burleigh LM, Steinhauer DA (2001) 
{\it Expert Reviews in Molecular Medicine} August 6, 1-18.

\bibitem{REV2} Shekel JJ, Wiley DC (2000) 
{\it Ann. Rev. Biochem.} 69:531-569.

\bibitem{400} Masaki I, Mizuno T, Kawasaki K (2006) 
{\it J. Biol. Chem.}  281:12729-12735.

\bibitem{HIVENDO0} Dimitrov DS (2004) 
{\it Nature Reviews} 2:109-122.

\bibitem{SENS} Sens P, Turner MS (2004)  
{\it Biophys. J.} 86:2049-2057.



\bibitem{DYNAMIN} Praefcke GJK, McMahon HT (2004) 
{\it Nature Reviews in Molecular Cell Biology} 5:133-147.

\bibitem{LOWPH} Duzgunes N, Pedroso de Lima MC, Stamatatos L,
Flasher D, Alford D, Friend DS, Nir S (1992) 
{\it J. Gen. Virology} 73:27-37.

\bibitem{MELIKYAN} Melikyan GB, Barnard RJO, Markosyan RM,
Young JAT, Cohen FS (2004) 
{\it J. Virol.} 78:3753-3762.

\bibitem{HARVARD} Mothes W, Boerger AL, Narayan S, Cunningham
JM, Young JAT (2000) 
{\it Cell} 103:679-689.

\bibitem{ANTIHARVARD} Diaz-Griffero F, Hoschander SA, 
Brojatsch J (2002) 
{\it J. Virol.}  76:12866-12876.

\bibitem{SFV} Marsh M, Bron R (1997)
{\it J. Cell Sci.}  110:95-103.

\bibitem{VACCINIA} Lai C, Gong S, Esteban M (1991) 
{\it J. Virology}  65:499-504.

\bibitem{HIVENDO1} Schaeffer E, Soros VB,  Greene WC (2004) 
{\it J. Virology}  78:1375-1383.

\bibitem{KABAT0} Platt EJ, Durnin JP, Kabat D (2005) 
{\it J. Virology}  79:4347-4356.

\bibitem{DCSIGN0} Su SV, Hong P, Baik S, Negrete OA,
Gurney KB,  Lee B (2004) 
{\it J. Biol. Chem.}  279:19122-19132.

\bibitem{DCSIGN1} Geijtenbeek TB, Kwon DS, Torensma R, van
Vliet SJ, van Duijnhoven GC, Middel J, Cornelissen IL,
Nottet HS, KewalRamani VN, Littman DR, Figdor CG, van Kooyk Y  (2000) 
{\it Cell}  100:587-597.

\bibitem{PLATELET} Chaipan C, Soilleux EJ, Simpson P, Hofmann
H, Gramberg T, Marzi A, Geier M, Stewart EA, Eisemann J,
Steinkasserer A, Suzuki-Inoue K, Fuller GL, Pearce AC,
Watson SP, Hoxie JA, Baribaud F, Pohlmann S (2006)
{\it J. Virology}  80:8951-8960. 

\bibitem{INF} Tumpey TM, {\it et al.} (2007) 
{\it Science} 315:655 - 659.

\bibitem{KABAT1} Kuhmann SE, Platt EJ, Kozak SL, Kabat D (2000) 
{\it J. Virology}  74:7005-7015.

\bibitem{SPIKESEM} Zhu P, Liu J, Bess J, Chertova E, Lifson
JD, Gris\'{e} H, Ofek GA, Taylor KA, Roux KH (2006) 
{\it Nature}  441:847-852.

\bibitem{LIGAND} D'Orsogna MR, Chou T (2005) 
{\it Phys. Rev. Lett.} 95:170603.

\bibitem{DESERNO} Deserno M (2004) 
{\it Phys. Rev. E} 69:031903.


\bibitem{UIUC} Yoon TY, Okumus B, Zhang F, Shin YK, Ha T (2006) 
{\it PNAS}  103:19731-19736.

\bibitem{TAS0} Melikyan GB, Markosyan RM, Hemmati H,
Delmedico MK, Lambert DM, Cohen FS (2000) 
{\it J. Cell Biol.}  151:413-423.

\bibitem{TAS1} Henderson HI, Hope TJ (2006) 
{\it Virology J.}  3:36.

\bibitem{DEFENSIN} Leikina E, Delanoe-Ayari H, Melikov K, Cho
MS, Chen A, Waring AJ, Wang W, Xie Y, Loo JA, Lehrer
RI, Chernomordik LV (2005) 
{\it Nature Immunology}  6:995-1001.

\end{thebibliography}
\end{document}